\documentclass[prl,aps,showpacs, twocolumn]{revtex4-1}

\def\be{\begin{equation}}
\def\ee{\end{equation}}

\usepackage{graphicx}
\usepackage{graphics}
\usepackage{bm}
\usepackage{amsmath}

\begin{document}

\title{Giant Charge Relaxation Resistance in the Anderson Model}

\author{Michele Filippone$^1$}
\author{Karyn Le Hur$^2$}
\author{Christophe Mora$^1$}
\affiliation{$^1$~Laboratoire Pierre Aigrain, \'Ecole Normale
  Sup\'erieure, Universit\'e Paris 7 Diderot, 
CNRS; 24 rue Lhomond, 75005 Paris, France}
\affiliation{$^2$~Departments of Physics and Applied Physics, Yale University, New Haven, Connecticut 06520, USA}

\date{\today}
\begin{abstract}
We investigate the dynamical charge response of the Anderson model 
viewed as a quantum RC circuit.
Applying a low-energy effective Fermi liquid
theory, a generalized Korringa-Shiba formula is derived at zero temperature, 
and the charge relaxation resistance
is expressed solely in terms of static susceptibilities which are accessible by
 Bethe ansatz.
We identify a giant charge relaxation resistance at intermediate magnetic fields
related to the destruction of the Kondo singlet. The scaling properties of this peak
are computed analytically in the Kondo regime. We also show that the resistance peak 
fades away at the particle-hole symmetric point.
\end{abstract}

 \maketitle

In recent years, an experimental effort has been devoted to
manipulate and measure electrons in nanoconductors 
in real time~\cite{Schoelkopf}.
At frequencies in the GHz range and cryogenic temperatures, current
and noise measurements provide information on the quantum dynamics of
charge carriers. Experiments have followed essentially two
directions, by either using on-chip quantum 
detectors~\cite{aguado2000,*deblock2003,*onac2006,*gustavsson2007,*xue2009,*basset2010} or by directly 
measuring the current using low noise amplifiers~\cite{gabelli2004,*gabelli2008,*zakka2007,*zakka2010}.
In an original experiment, Gabelli et al.~\cite{gabelli2006} 
have realized the quantum equivalent of an RC circuit with a quantum dot connected to a 
spin-polarized single-lead reservoir and capacitively coupled to a metallic top gate,
this setup being later used as a single-electron source~\cite{feve2007,*mahe2010}.
By applying an AC modulation to the gate voltage,
they measured the admittance of the dot at low frequency.
A comparison with the classical RC circuit allows to extract
 a capacitance and a charge relaxation resistance. Their measurements have confirmed 
the prediction~\cite{buttiker1993,*buttiker1993b,nigg2006}
of a quantized charge relaxation resistance $R_q = h/2 e^2$ where $e$ is the electron charge
and $h$ the Planck constant.
This prediction was recently shown~\cite{mora2010,hamamoto2010} to be valid for all interaction
strength. Ref.~\cite{mora2010} also predicted the emergence of an additional
universal resistance $R_q = h/e^2$ in the case of a large dot.

The quantum RC circuit is described by the Anderson
model~\cite{hewson1997} 
when the level spacing is sufficiently large and electron transport is not
spin-polarized in contrast to 
Refs.~\cite{buttiker1993,*buttiker1993b,nigg2006,mora2010,hamamoto2010}. 
In that case, the gate
voltage controls the dot single-particle energy $\varepsilon_d (<0)$. 
In addition to being experimentally relevant, for the transport
through short nanotubes~\cite{delattre2009}, 
small quantum dots or even molecules~\cite{roch2008},
the Anderson model is fascinating because it displays features
of strong correlation with the emergence of Kondo physics at low energy.
The question of how these correlations affect the quantization of the 
charge relaxation resistance is a fundamental issue.

 The linear charge response of the quantum dot 
to a gate voltage oscillation 
defines the capacitance $C_0$ and the resistance $R_q$
via the low frequency expansion~\cite{mora2010}
\begin{equation}\label{expansion}
\frac{e^2 \langle \hat{n} (\omega) \rangle}
{- \varepsilon_d (\omega)} = C_0 + i \omega C_0^2 R_q
+ {\cal O} (\omega^2)
\end{equation}
where $\hat{n}$ the number of electrons
on the dot.
The capacitance is thus the static response of the dot.
The product $\omega C_0^2 R_q$ describes relaxation towards the 
changing ground state that implies energy 
dissipation~\cite{mora2010,rodionov2009}.
In this Letter, we investigate the dynamical charge response of the Anderson model at zero temperature and finite magnetic fields and we evidence a giant charge relaxation resistance phenomenon associated with the destruction of the Kondo effect at intermediate fields. 

More precisely, by applying a low-energy effective Fermi liquid
theory~\cite{nozieres1974,krishna1980a,*krishna1980b}, 
we derive a generalized Korringa-Shiba formula~\cite{shiba1975} for the charge
susceptibility that extends to finite magnetic fields. The charge relaxation resistance
then depends only on {\it static} susceptibilities that are computed 
analytically
resorting to the Bethe ansatz solution~\cite{tsvelick1983,kawakami1982,*okiji1982,hewson1997} 
in the Kondo regime.
At zero magnetic field, the original Korringa-Shiba~\cite{shiba1975} formula predicts the quantized value  $R_q = h/4 e^2$ independent of interactions.  This result agrees with the non-interacting scattering approach with two (spin) conducting  channels~\cite{buttiker1993,*buttiker1993b,nigg2006}. At  large magnetic fields, the dot becomes spin-polarized, reducing electron transfer
in both spin channels, and the quantized value $R_q = h/4 e^2$ is recovered perturbatively.
 In the crossover regime between these two limits, a peak was observed in the Numerical Renormalization Group (NRG) calculations of Ref.~\cite{lee2011}, where it is attributed to spin fluctuations in the dot. Hereafter, we derive analytically the emergence of this peak in the Kondo regime and derive its scaling properties. In particular, the peak is found to disappear completely at the particle-hole symmetric point.

The origin of the peak 
in the resistance is related to the destruction of the Kondo
singlet by the magnetic field 
which gives more flexibility to the spin configuration, while
the charge remains frozen by interactions. As a result, a change
in the gate voltage significantly modifies the magnetization,
with an increase of dissipation by particle-hole excitations.
The charge however is relatively insensitive to the gate
voltage and the capacitance remains small. An increasing
dissipation $\propto C_0^2 R_q$ with an almost constant capacitance
$C_0$ thus leads to an increasing charge relaxation resistance $R_q$.
A further increase of the magnetic field eventually 
polarizes the spin on the dot, reduces spin flexibility and
thereby energy dissipation. Hence, the charge relaxation 
resistance $R_q$ passes through
a maximum when the Zeeman energy is comparable to the Kondo energy.

The Hamiltonian of the Anderson model is given by
\begin{equation}\label{am}
\begin{split}
H &= \sum_{\sigma,k} \varepsilon_{k\sigma} 
 c^\dagger_{k\sigma} c_{k\sigma}
+ \sum_\sigma \varepsilon_{d\sigma} 
\, \hat{n}_\sigma \\
& + U \hat{n}_{\uparrow} \hat{n}_{\downarrow}
+ t \sum_{k,\sigma} \left( c_{k\sigma}^\dagger  d_\sigma +  d_\sigma^\dagger
c_{k\sigma} \right),
\end{split}
\end{equation}
with $\hat{n}_\sigma = d_\sigma^\dagger  d_\sigma$ the number of spin $\sigma$
electron on the dot, $\hat{n} = \hat{n}_\uparrow + \hat{n}_\downarrow$,
the linear  spectrum $\varepsilon_{k\sigma}
= \varepsilon_k - g\sigma\mu_B B/2$ of conduction electrons
characterized by the constant density of states $\nu_0$,
and the energy levels $\varepsilon_{d\sigma}
= \varepsilon_d - g\sigma\mu_B B/2$ of the dot.  $\mu_B$
is the Bohr magneton, $g$ the Lande factor, $B$ the external magnetic field and $\sigma=\pm$ refers to $\uparrow$, $\downarrow$ states, respectively. The two terms in the second line of Eq.~\eqref{am} describe
respectively Coulomb interaction and tunneling from the dot to the lead with the hybridization constant $\Gamma = \pi \nu_0 t^2$.
In the presence of an AC drive of very small amplitude, 
$\varepsilon_d = \varepsilon_d^0 +
\varepsilon_\omega \cos \omega t$ with $\varepsilon_\omega \to 0$,
the system relaxes towards the evolving ground state of the Hamiltonian
and the dissipated power
\begin{equation} \label{dissip}
{\cal P} = \frac{1}{2} \varepsilon_\omega^2 \, \omega 
\, {\rm Im} \chi_c (\omega),
\end{equation}
which is given by linear response theory,  is proportional to the imaginary part of the dynamical
charge susceptibility $\chi_c (t-t') = i \theta (t-t') \langle
[ \hat{n} (t), \hat{n} (t') ] \rangle$.

NRG calculations~\cite{krishna1980a,*krishna1980b} 
and RG arguments~\cite{haldane1978} have shown
that the low energy properties of the Anderson model~\eqref{am}
are always those of a Fermi liquid. The effective Fermi liquid
Hamiltonian takes the form
\begin{equation}\label{fermi}
H = \sum_{\sigma,k} \varepsilon_{k\sigma}  a^\dagger_{k\sigma} a_{k\sigma}
+  \sum_{k,k', \sigma} K_\sigma (\varepsilon_d) 
a^\dagger_{k\sigma} a_{k'\sigma}.
\end{equation}
The free quasiparticles of the first term are related to the original fermions
$c_{k\sigma}$ by a phase shift of $\pi/2$.
The second term is a marginal perturbation
corresponding to a potential scattering at the impurity site.
It defines a line of fixed points parametrized by $\varepsilon_d$
connecting the Kondo regime (for $\varepsilon_d \simeq -U/2$) to
the mixed valence regimes (for $\varepsilon_d \simeq 0$ or
$\varepsilon_d \simeq -U$). The potential is related to the mean
occupation of the dot via the Friedel sum rule
$\langle \hat{n}_\sigma \rangle = 1/2 - \frac{1}{\pi} 
\arctan \left[ \pi \nu_0
K_\sigma (\varepsilon_d ) \right]$. Note that the potentials $K_\sigma (\varepsilon_d)$ formally also depend
on $U$, $\Gamma$ and $B$. Again we study the response to
the AC drive  $\varepsilon_d = \varepsilon_d^0 +
\varepsilon_\omega \cos \omega t$ with $\varepsilon_\omega, \omega \to 0$.
Expanding the potentials as $K_\sigma (\varepsilon_d) = K_\sigma^0 
+ K_\sigma' (\varepsilon_{d}^0)  \, \varepsilon_\omega \cos \omega t$,
we change the basis
 to the one-particle states~\cite{krishna1980a,*krishna1980b} that diagonalize
the potential scattering terms $K_\sigma^0 = K_\sigma (\varepsilon_d^0)$.
The remaining scattering term in the Hamiltonian is given by
\begin{equation}\label{potsca}
\varepsilon_\omega \cos \omega t 
\sum_\sigma \frac{K_\sigma' (\varepsilon_{d}^0)}{1+(\pi \nu_0
  K_\sigma^0)^2}
\, \sum_{k,k'} \tilde{a}^\dagger_{k\sigma} \tilde{a}_{k'\sigma},
\end{equation}
with the new  quasiparticles $\tilde{a}_{k'\sigma}$. 
The derivative of the occupation numbers with 
respect to $\varepsilon_d$ in the Friedel sum rule formula above
introduces the static spin-dependent susceptibilities 
$\chi_{c\sigma} = - \partial \langle \hat{n}_\sigma
\rangle / \partial \varepsilon_d$. Once inserted into
Eq.~\eqref{potsca}, we obtain
\begin{equation}\label{lowener2}
H = \sum_{\sigma,k} \varepsilon_{k\sigma}  
\tilde{a}^\dagger_{k\sigma} \tilde{a}_{k\sigma}
+ \varepsilon_\omega 
\cos \omega t \,
 \sum_\sigma \frac{\chi_{c\sigma}}{\nu_0} \,  
\sum_{k,k'}  \tilde{a}^\dagger_{k\sigma} \tilde{a}_{k'\sigma}.
\end{equation}
In the static case $\omega=0$, the second term in Eq.~\eqref{lowener2}
adds the phase shift 
$\delta_\sigma = - \pi \nu_0 \varepsilon_\omega\chi_{c\sigma}/\nu_0$. 
The Friedel sum rule translates it into a shift in the
occupations $\delta \langle \hat{n}_\sigma \rangle =  - \chi_{c\sigma}
 \, \varepsilon_\omega$
in agreement with the definition of the charge susceptibilities.
The Hamiltonian in Eq.~\eqref{lowener2}
is extremely general and it only assumes
a low-energy Fermi liquid fixed point. A similar model can be found
in Ref.~\cite{garst2005} where the spin susceptibility is discussed.

Interestingly, the low energy model  Eq.~\eqref{lowener2} provides an alternative
to compute the dissipated power Eq.~\eqref{dissip}. 
Following standard linear response theory, it involves the  operators 
$\hat{A}_\sigma = (\chi_{c\sigma} /\nu_0) 
\sum_{k,k'} \tilde{a}^\dagger_{k\sigma} \tilde{a}_{k'\sigma}$,
coupled to the AC excitation in Eq.~\eqref{lowener2},
namely
\begin{equation} \label{dissip2}
{\cal P} = \frac{1}{2} \varepsilon_\omega^2 \, \omega 
\, \sum_\sigma {\rm Im}  \chi_{\hat{A}\sigma} (\omega),
\end{equation}
with  $\chi_{\hat{A}\sigma} (t-t') = i \theta (t-t') \langle
[ \hat{A}_\sigma (t), \hat{A}_\sigma (t') ] \rangle$.
The operators $\hat{A}_\sigma$ create particle-hole pairs 
that are responsible for energy dissipation.
 The calculation is straightforward
and gives, at zero temperature, 
${\rm Im}  \chi_{\hat{A}\sigma} (\omega) = \pi \chi_{c\sigma}^2
\omega$, {\it i.e.}, proportional to the density of available particle-hole
pairs with energy $\omega$.
An identification of Eqs.~\eqref{dissip} and~\eqref{dissip2}
finally results in our generalized Korringa-Shiba formula
\begin{equation}\label{korringa}
{\rm Im} \chi_c (\omega) = 
\pi \omega \left( \chi_{c\uparrow}^2 + \chi_{c\downarrow}^2 
\right),
\end{equation}
obtained to lowest order~\footnote{
The effective Hamiltonian in Eq.~\eqref{fermi}
is usually accompanied by the Fermi liquid corrections
introduced by Nozi\` eres~\cite{nozieres1974,lesage1999,*mora2009}.
These corrections correspond however to irrelevant operators and give only
subleading contributions to the generalized Korringa-Shiba
Eq.~\eqref{korringa}.}
 in $\omega$. 
The physical meaning
of this expression is explicit. In the presence of
the AC driving applied to the gate voltage, 
relaxation is necessary to adjust
the occupation numbers to the instant ground state of the 
Hamiltonian. This relaxation is realized
 by particle-hole excitations, in each
spin sector independently, with 
amplitudes (see Eq.~\eqref{lowener2})
 that are determined by the static charge susceptibilities
$\chi_{c\sigma}$
controlling the variations of the spin populations with the gate voltage.
Eq.~\eqref{korringa} simply states that the energy dissipated 
in the relaxation mechanism increases quadratically with these
amplitudes as a result of the Fermi golden rule.

At a general level, the low frequency properties of the quantum RC circuit
characterized by Eq.~\eqref{expansion}
derive from the knowledge of the dynamical charge susceptibility
since $\langle \hat{n} (\omega) \rangle = - \chi_c (\omega) \varepsilon_d (\omega)$
in the linear regime.
The  capacitance $C_0 = e^2 \chi_c$ is solely determined by the
static charge susceptibility $\chi_c = \chi_c (\omega \to0)
= \chi_{c\uparrow} + \chi_{c\downarrow} $ that is calculated using
Bethe ansatz. Hence, measuring the capacitance 
realizes a charge spectroscopy~\cite{LeHur,*cottet2011}.
At zero magnetic field and large enough interaction, $U > \Gamma$,
$\chi_c$ develops a double-peak structure as a function of $\varepsilon_d$:
maximum in the valence regimes with a valley in the
intermediate Kondo regime around  $\varepsilon_d = -U/2$ where 
$\chi_c = 8 \Gamma / \pi U^2$ for  $U \gg \Gamma$.
This strong reduction of capacitance, or charge sensitivity,  characterizes the Kondo limit
where the charge  is frozen.
It contradicts the non-interacting scattering 
theory~\cite{gabelli2006,buttiker1993,*buttiker1993b} where the capacitance is
proportional to the density of state and would therefore reveal the Kondo resonance~\cite{lee2011}.
The two approaches are nonetheless reconciled by noting that, the Kondo
resonance is mostly exhausted by spin fluctuations, and the density of states
of charge excitations of the Anderson model, the holons, reproduces~\cite{kawakami1990}
the exact value of the charge susceptibility and thus of the
capacitance.

%These results may seem counter-intuitive if one recalls that capacitance and 
%density of states are proportional in the non-interacting 
%scattering theory~\cite{gabelli2006,buttiker1993,*buttiker1993b} 
%and that the Kondo resonance is characterized by a peak in the
%density of states. The reason is that the Kondo resonance is almost
%completely exhausted by spin fluctuations while the capacitance is only sensitive
%to charge fluctuations. Experimentally, a capacitance spectroscopy would 
%therefore not reveal the Kondo resonance.
%In fact, the density of states for the elementary 
%charge excitations of the Anderson model, the holons, has been computed in Ref.~\cite{kawakami1990}:
%it reproduces the exact value of the charge susceptibility and of the
%capacitance. The non-interacting expression of the quantum 
%capacitance~\cite{buttiker1993,*buttiker1993b} is recovered when only charge excitations are taken into account.
The Korringa-Shiba Eq.~\eqref{korringa} substituted in the expansion 
Eq.~\eqref{expansion} expresses the resistance $R_q$ in terms of
static susceptibilities,
computable by  Bethe ansatz.
At zero magnetic field, $\chi_{c\uparrow} = \chi_{c\downarrow} 
= \chi_c/2$, Eq.~\eqref{korringa} reproduces the standard Korringa-Shiba
relation
% In the simplest case of a vanishing magnetic field, the 
%spin-SU(2) symmetry is preserved and $\chi_{c\uparrow} = \chi_{c\downarrow} 
%= \chi_c/2$. 
and the charge relaxation resistance is  found to be quantized
and universal, 
\begin{equation}\label{quantized}
R_q = \frac{h}{4 e^2},
\end{equation}
in agreement with the scattering approach involving 
two equivalent spin channels~\cite{buttiker1993,*buttiker1993b,nigg2006}.
In the general case, we introduce the {\it charge magneto-susceptibility} 
$\chi_m = \chi_{c\uparrow} - \chi_{c\downarrow}$ which measures the sensitivity
of the magnetization to a change in the gate voltage. The resistance
reads
\begin{equation}\label{resistance}
R_q = \frac{h}{4 e^2} \, \frac{\chi_c^2 + \chi_m^2}{\chi_c^2}.
\end{equation}
For $\varepsilon_d = -U/2$, particle-hole symmetry implies that the
magnetization is extremal with respect to the gate voltage and
$\chi_m$ identically vanishes.
Eq.~\eqref{quantized} is thus obtained for  all ratios of $U/\Gamma$.

% The resistance $R_q$ is thus flat in the intermediate magnetic field regime, equal to the quantized value
%Eq.~\eqref{quantized} for all ratios of $U/\Gamma$.

In the rest of this Letter, we focus on the Kondo regime $U \gg \Gamma$ 
where the gate voltage explores the valley between the
Coulomb peaks located around $\varepsilon_d \simeq 0$ and $\varepsilon_d \simeq -U$.
Far enough from these 
Coulomb peaks, $|\varepsilon_d| /\Gamma \gg \ln ( U/\Gamma )$, 
the charge on the dot remains of order one and the 
renormalization~\cite{haldane1978} of the
peak positions is negligible. The charge susceptibility is computed perturbatively at zero magnetic field,
\begin{equation}\label{chargesus}
\chi_c = \frac{\Gamma}{\pi} \left( \frac{1}{(\varepsilon_d +U)^2}
+  \frac{1}{\varepsilon_d^2} \right),
\end{equation}
and remains constant as the magnetic field is increased with $g \mu_B B
\ll \sqrt{|\varepsilon_d| \Gamma}$. 

\begin{figure}
\centerline{\includegraphics[width=\columnwidth]{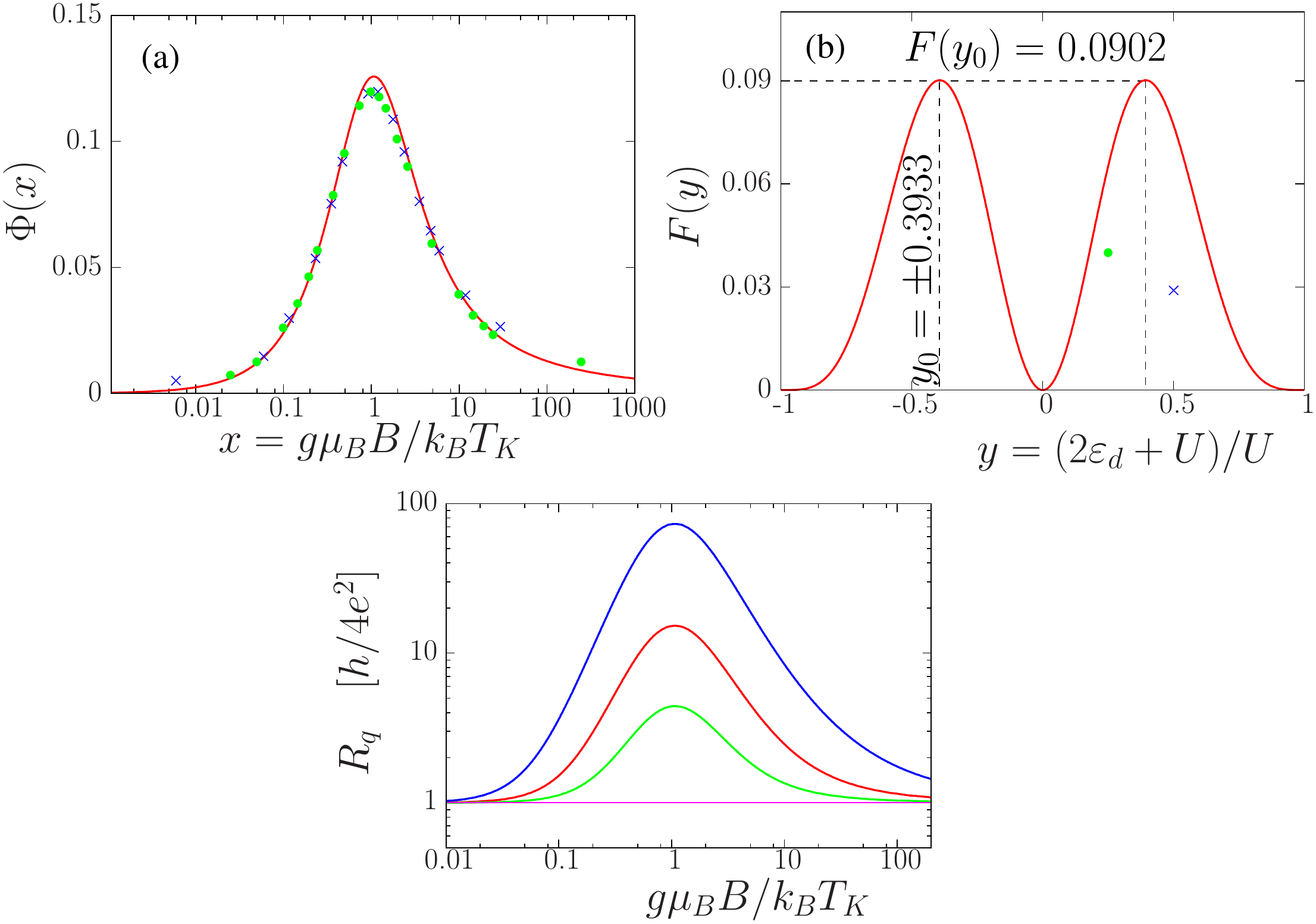}}
%\centerline{\includegraphics[width=0.7\columnwidth]{functionphi-mod.pdf}}
%\centerline{\includegraphics[width=0.7\columnwidth]{fonctiongrandF-mod.pdf}}
\caption{(a) Scaling function $\Phi(x)$ (full line) and
 (b) Envelope function $F(y)$ (see main text). 
$F(0)=0$ at the particle-hole symmetric point $y=0$ where
 $\partial T_K / \partial \varepsilon_d = 0$.  
$y=\pm1$ correspond to
the two Coulomb peaks in transport or valence regime.
Green circles ($\varepsilon_d=-0.15$) and blue crosses ($\varepsilon_d=-0.1$)
are extracted from Fig. 3a) of Ref.~\cite{lee2011}, with $U=0.4$
and $\Gamma=0.02$, by implementing Eq.~\eqref{scaling2} and
rescaling the $x$ axis.
(c) Charge relaxation resistance for $\varepsilon_d/U = -1/2\pm 0.1967$ and
various ratios of $U/\Gamma=15,10,7$.}
\label{all-figs}
\end{figure}
For moderate magnetic fields $g \mu_B B\ll \sqrt{|\varepsilon_d| \Gamma}$, the magnetization of the dot is known~\cite{tsvelick1983,hewson1997} from the Bethe ansatz solution of the Anderson model. In the Kondo limit, it exhibits the scaling form,
\begin{equation}\label{magnetization}
m = \frac{\langle \hat{n}_\uparrow \rangle
- \langle \hat{n}_\downarrow \rangle }{2}
= f \left(\frac{g \mu_B B}{k_B T_K} \right),
\end{equation}
where  $T_K = 2 \sqrt{U \Gamma/\pi {\rm e}}\exp[{\pi \varepsilon_d(\varepsilon_d +U)/2 U \Gamma}]$ is the Kondo temperature and the scaling function $f(x)$ connects the asymptotes 
$f(x) = x/\sqrt{2 \pi {\rm e}}$ for $x\ll 1$, and $f(x) = 1/2-1/(4 \ln x)$
for $x \gg 1$, {\it i.e.}, for low and large magnetic fields,  ${\rm e}$ 
referring to Euler's number.
The dependence of the magnetization Eq.~\eqref{magnetization} on
the gate voltage, or $\varepsilon_d$, is via the Kondo
temperature. Computing the derivative of the Kondo temperature
with respect to $\varepsilon_d$, then one finds
\begin{equation}\label{scaling}
\chi_m = \frac{\pi}{ \Gamma} \, 
\frac{2 \varepsilon_d +U}{U} \, 
\Phi \left( \frac{g \mu_B B}{k_B T_K} \right).
\end{equation}
The charge magneto-susceptibility $\chi_m$ is a an odd 
function of $\varepsilon_d +U/2$
that vanishes at the particle-hole symmetric point.
The scaling function $\Phi (x)= x f'(x)$ is represented Fig.~\ref{all-figs}(a)
in good agreement with Ref.~\cite{lee2011}.
 It exhibits a peak at $x_0 = 1.0697$ with $\Phi(x_0) =  0.1257$.
% and has asymptotes
%\begin{equation}
%\begin{split}
%\Phi (x) = \begin{cases} \displaystyle
%\frac{x}{\sqrt{2 \pi e}}, \qquad \quad {\rm at} \, \, x \ll 1,  \\[3mm] \displaystyle
%\frac{1}{4} \, \frac{1}{(\ln x)^2}, \qquad {\rm at} \, \, x \gg 1.
%\end{cases}
%\end{split}
%\end{equation}
Inserting the results Eqs.~\eqref{chargesus} and~\eqref{scaling} into 
Eq.~\eqref{resistance}, the scaling form of the charge relaxation
resistance is obtained,
\begin{equation}\label{scaling2}
R_q = \frac{h}{4 e^2}  \left[ 1+ \left(\frac{U}{\Gamma}\right)^4 
F (y) \left(\Phi (x)\right)^2
\right],
\end{equation}
with $y=(2 \varepsilon_d+U)/U$, $x=g \mu_B B/k_B T_K$. 
The peak
in the 
resistance as a function of the magnetic field is described
by the scaling function $\Phi(x)$.
It is also  weighted by the envelope
$F(y) = (\pi^2/8)^2 y^2 (y^2-1)^4/(1+y^2)^2$, shown Fig.~\ref{all-figs}(b).
The agreement with Ref.~\cite{lee2011}, where $U/\Gamma=20$
is finite, is here only approximate.
%, that has maxima for $y_0 = \pm 0.3933$ with $F(y_0) = 0.09016$.
The global maximum in the resistance is thus 
obtained for $\varepsilon_d/U = -1/2\pm 0.1967$, 
$g \mu_B B = 1.0697\, k_B T_K$, with 
$R_q = 0.00142 \, (h/4 e^2) (U/\Gamma)^4$ 
which predicts a strong increase of the resistance maximum with the ratio $U/\Gamma$,
as seen Fig.~\ref{all-figs}(c). 

For large magnetic fields $g \mu_B B \gg \sqrt{|\varepsilon_d| \Gamma}$, the free
orbital regime~\cite{haldane1978} is reached and 
straightforward perturbation theory applies.
The result is
\[
\langle \hat{n}_\uparrow \rangle = 1 - \frac{1}{\pi}\frac{\Gamma}{\varepsilon_M -\varepsilon_d},
\quad
\langle \hat{n}_\downarrow \rangle = \frac{1}{\pi}\frac{\Gamma}{\varepsilon_M+\varepsilon_d+U},
\]
with $\varepsilon_M = g \mu_B B/2$. 
This leads to $\chi_m=0$ for $\varepsilon_d = -U/2$ as expected and, 
for very large magnetic fields $g \mu_B B \gg (|\varepsilon_d|, \varepsilon_d+U)$, 
the quantized value Eq.~\eqref{quantized} is recovered for all gate voltages.
Note that the standard 
result~\cite{buttiker1993,*buttiker1993b,nigg2006,mora2010,hamamoto2010} $R_q=h/2 e^2$
is only recovered for a fully polarized Fermi sea in the lead.

To summarize, the peak in the charge relaxation resistance is due to the enhancement of
$\chi_m$ while the total charge remains quenched and $\chi_c$ small.
In the presence of a finite magnetic field, the Kondo state is a mixture
of singlet and triplet spin configurations controlled by the ratio of the
Zeeman energy to the Kondo energy. A change in the gate voltage modifies this
ratio and, while keeping the total charge of the dot almost constant, redistributes
the spin up and down occupations. This leads to a larger number of particle-hole
excitations for each spin species and therefore increases dissipation.
%In the case of zero or very large magnetic field, the spin configuration is frozen
%and no spin redistribution occurs by changing the gate voltage. The resistance is 
%then of order one in units of $h/e^2$. 
At the particle-hole symmetric point, the Kondo
energy is stationary with respect to the gate voltage such that 
no spin redistribution occurs and the peak in the resistance is absent. It is worth mentioning that
the predicted peak in the charge relaxation resistance occurring at intermediate magnetic fields
can be observed using current technology \cite{gabelli2006}. 
This work can be extended in various directions, by considering either Zeeman effects on a large cavity characterized by several energy levels \cite{KLH} or a large number of channels \cite{etzioni2011}. 
We finally stress that our result~\eqref{resistance} for the resistance is valid not
only in the Kondo regime but for all values of $U$, $\varepsilon_d$ and $B$.

We acknowledge discussions with A. Cottet, T. Kontos, M. Lee, H. Baranger, G. Finkelstein.
K.L.H. was supported by Department of Energy, under the grant DE-FG02-08ER46541, and by the Yale Center
for Quantum Information Physics (NSF DMR-0653377).

\bibliography{bibliographie}

%Merlin.mbs v4.21 2009-07-09.
\begin{thebibliography}{10}%
\makeatletter
\providecommand \@ifxundefined [1]{%
 \ifx #1\undefined \expandafter \@firstoftwo
 \else \expandafter \@secondoftwo
\fi
}%
\providecommand \@ifnum [1]{%
 \ifnum #1\expandafter \@firstoftwo
 \else \expandafter \@secondoftwo
\fi
}%
\providecommand \enquote [1]{``#1''}%
\providecommand \bibnamefont  [1]{#1}%
\providecommand \bibfnamefont [1]{#1}%
\providecommand \citenamefont [1]{#1}%
\providecommand\href[0]{\@sanitize\@href}%
\providecommand\@href[1]{\endgroup\@@startlink{#1}\endgroup\@@href}%
\providecommand\@@href[1]{#1\@@endlink}%
\providecommand \@sanitize [0]{\begingroup\catcode`\&12\catcode`\#12\relax}%
\@ifxundefined \pdfoutput {\@firstoftwo}{%
 \@ifnum{\z@=\pdfoutput}{\@firstoftwo}{\@secondoftwo}%
}{%
 \providecommand\@@startlink[1]{\leavevmode\special{html:<a href="#1">}}%
 \providecommand\@@endlink[0]{\special{html:</a>}}%
}{%
 \providecommand\@@startlink[1]{%
  \leavevmode
  \pdfstartlink
   attr{/Border[0 0 1 ]/H/I/C[0 1 1]}%
   user{/Subtype/Link/A<</Type/Action/S/URI/URI(#1)>>}%
  \relax
 }%
 \providecommand\@@endlink[0]{\pdfendlink}%
}%
\providecommand \url  [0]{\begingroup\@sanitize \@url }%
\providecommand \@url [1]{\endgroup\@href {#1}{\urlprefix}}%
\providecommand \urlprefix [0]{URL }%
\providecommand \Eprint[0]{\href }%
\@ifxundefined \urlstyle {%
  \providecommand \doi [1]{doi:\discretionary{}{}{}#1}%
}{%
  \providecommand \doi [0]{doi:\discretionary{}{}{}\begingroup
  \urlstyle{rm}\Url }%
}%
\providecommand \doibase [0]{http://dx.doi.org/}%
\providecommand \Doi[1]{\href{\doibase#1}}%
\providecommand \bibAnnote [3]{%
  \BibitemShut{#1}%
  \begin{quotation}\noindent
    \textsc{Key:}\ #2\\\textsc{Annotation:}\ #3%
  \end{quotation}%
}%
\providecommand \bibAnnoteFile [2]{%
  \IfFileExists{#2}{\bibAnnote {#1} {#2} {\input{#2}}}{}%
}%
\providecommand \typeout [0]{\immediate \write \m@ne }%
\providecommand \selectlanguage [0]{\@gobble}%
\providecommand \bibinfo [0]{\@secondoftwo}%
\providecommand \bibfield [0]{\@secondoftwo}%
\providecommand \translation [1]{[#1]}%
\providecommand \BibitemOpen[0]{}%
\providecommand \bibitemStop [0]{}%
\providecommand \bibitemNoStop [0]{.\EOS\space}%
\providecommand \EOS [0]{\spacefactor3000\relax}%
\providecommand \BibitemShut [1]{\csname bibitem#1\endcsname}%
%</preamble>
\bibitem{Schoelkopf}%
  \BibitemOpen
  \bibfield{author}{%
  \bibinfo {author} {\bibfnamefont{R.~J.}\ \bibnamefont{Schoelkopf}}, \bibinfo
  {author} {\bibfnamefont{P.}~\bibnamefont{Wahlgren}}, \bibinfo {author}
  {\bibfnamefont{A.~A.}\ \bibnamefont{Kozhevnikov}}, \bibinfo {author}
  {\bibfnamefont{P.}~\bibnamefont{Delsing}},\ and\ \bibinfo {author}
  {\bibfnamefont{D.~E.}\ \bibnamefont{Prober}},\ }%
  \bibfield{journal}{%
  \bibinfo {journal} {Science}\ }%
  \textbf{\bibinfo {volume} {280}},\ \bibinfo {pages} {1238} (\bibinfo {year}
  {1998})%
  \bibAnnoteFile{NoStop}{Schoelkopf}%
\bibitem{aguado2000}%
  \BibitemOpen
  \bibfield{author}{%
  \bibinfo {author} {\bibfnamefont{R.}~\bibnamefont{Aguado}}\ and\ \bibinfo
  {author} {\bibfnamefont{L.~P.}\ \bibnamefont{Kouwenhoven}},\ }%
  \bibfield{journal}{%
  \Doi{10.1103/PhysRevLett.84.1986}{\bibinfo {journal} {Phys. Rev. Lett.}}\ }%
  \textbf{\bibinfo {volume} {84}},\ \bibinfo {pages} {1986} (\bibinfo {year}
  {2000})%
  \bibAnnoteFile{NoStop}{aguado2000}%
\bibitem{deblock2003}%
  \BibitemOpen
  \bibfield{author}{%
  \bibinfo {author} {\bibfnamefont{R.}~\bibnamefont{Deblock}} \emph{et~al.},\
  }%
  \bibfield{journal}{%
  \Doi{10.1126/science.1084175}{\bibinfo {journal} {Science}}\ }%
  \textbf{\bibinfo {volume} {301}},\ \bibinfo {pages} {203} (\bibinfo {year}
  {2003})%
  \bibAnnoteFile{NoStop}{deblock2003}%
\bibitem{onac2006}%
  \BibitemOpen
  \bibfield{author}{%
  \bibinfo {author} {\bibfnamefont{E.}~\bibnamefont{Onac}} \emph{et~al.},\ }%
  \bibfield{journal}{%
  \Doi{10.1103/PhysRevLett.96.176601}{\bibinfo {journal} {Phys. Rev. Lett.}}\
  }%
  \textbf{\bibinfo {volume} {96}},\ \bibinfo {pages} {176601} (\bibinfo {year}
  {2006})%
  \bibAnnoteFile{NoStop}{onac2006}%
\bibitem{gustavsson2007}%
  \BibitemOpen
  \bibfield{author}{%
  \bibinfo {author} {\bibfnamefont{S.}~\bibnamefont{Gustavsson}}
  \emph{et~al.},\ }%
  \bibfield{journal}{%
  \Doi{10.1103/PhysRevLett.99.206804}{\bibinfo {journal} {Phys. Rev. Lett.}}\
  }%
  \textbf{\bibinfo {volume} {99}},\ \bibinfo {pages} {206804} (\bibinfo {year}
  {2007})%
  \bibAnnoteFile{NoStop}{gustavsson2007}%
\bibitem{xue2009}%
  \BibitemOpen
  \bibfield{author}{%
  \bibinfo {author} {\bibfnamefont{W.}~\bibnamefont{Xue}} \emph{et~al.},\ }%
  \bibfield{journal}{%
  \bibinfo {journal} {Nature Physics}\ }%
  \textbf{\bibinfo {volume} {5}},\ \bibinfo {pages} {660} (\bibinfo {year}
  {2009})%
  \bibAnnoteFile{NoStop}{xue2009}%
\bibitem{basset2010}%
  \BibitemOpen
  \bibfield{author}{%
  \bibinfo {author} {\bibfnamefont{J.}~\bibnamefont{Basset}}, \bibinfo {author}
  {\bibfnamefont{H.}~\bibnamefont{Bouchiat}},\ and\ \bibinfo {author}
  {\bibfnamefont{R.}~\bibnamefont{Deblock}},\ }%
  \bibfield{journal}{%
  \Doi{10.1103/PhysRevLett.105.166801}{\bibinfo {journal} {Phys. Rev. Lett.}}\
  }%
  \textbf{\bibinfo {volume} {105}},\ \bibinfo {pages} {166801} (\bibinfo {year}
  {2010})%
  \bibAnnoteFile{NoStop}{basset2010}%
\bibitem{gabelli2004}%
  \BibitemOpen
  \bibfield{author}{%
  \bibinfo {author} {\bibfnamefont{J.}~\bibnamefont{Gabelli}} \emph{et~al.},\
  }%
  \bibfield{journal}{%
  \Doi{10.1103/PhysRevLett.93.056801}{\bibinfo {journal} {Phys. Rev. Lett.}}\
  }%
  \textbf{\bibinfo {volume} {93}},\ \bibinfo {pages} {056801} (\bibinfo {year}
  {2004})%
  \bibAnnoteFile{NoStop}{gabelli2004}%
\bibitem{gabelli2008}%
  \BibitemOpen
  \bibfield{author}{%
  \bibinfo {author} {\bibfnamefont{J.}~\bibnamefont{Gabelli}}\ and\ \bibinfo
  {author} {\bibfnamefont{B.}~\bibnamefont{Reulet}},\ }%
  \bibfield{journal}{%
  \Doi{10.1103/PhysRevLett.100.026601}{\bibinfo {journal} {Phys. Rev. Lett.}}\
  }%
  \textbf{\bibinfo {volume} {100}},\ \bibinfo {pages} {026601} (\bibinfo {year}
  {2008})%
  \bibAnnoteFile{NoStop}{gabelli2008}%
\bibitem{zakka2007}%
  \BibitemOpen
  \bibfield{author}{%
  \bibinfo {author} {\bibfnamefont{E.}~\bibnamefont{Zakka-Bajjani}}
  \emph{et~al.},\ }%
  \bibfield{journal}{%
  \Doi{10.1103/PhysRevLett.99.236803}{\bibinfo {journal} {Phys. Rev. Lett.}}\
  }%
  \textbf{\bibinfo {volume} {99}},\ \bibinfo {pages} {236803} (\bibinfo {year}
  {2007})%
  \bibAnnoteFile{NoStop}{zakka2007}%
\bibitem{zakka2010}%
  \BibitemOpen
  \bibfield{author}{%
  \bibinfo {author} {\bibfnamefont{E.}~\bibnamefont{Zakka-Bajjani}}
  \emph{et~al.},\ }%
  \bibfield{journal}{%
  \Doi{10.1103/PhysRevLett.104.206802}{\bibinfo {journal} {Phys. Rev. Lett.}}\
  }%
  \textbf{\bibinfo {volume} {104}},\ \bibinfo {pages} {206802} (\bibinfo {year}
  {2010})%
  \bibAnnoteFile{NoStop}{zakka2010}%
\bibitem{gabelli2006}%
  \BibitemOpen
  \bibfield{author}{%
  \bibinfo {author} {\bibfnamefont{J.}~\bibnamefont{Gabelli}}, \bibinfo
  {author} {\bibfnamefont{G.}~\bibnamefont{F\`eve}}, \bibinfo {author}
  {\bibfnamefont{J.-M.}\ \bibnamefont{Berroir}}, \bibinfo {author}
  {\bibfnamefont{B.}~\bibnamefont{Pla{\c{c}}ais}}, \bibinfo {author}
  {\bibfnamefont{A.}~\bibnamefont{Cavanna}}, \bibinfo {author}
  {\bibfnamefont{B.}~\bibnamefont{Etienne}}, \bibinfo {author}
  {\bibfnamefont{Y.}~\bibnamefont{Jin}},\ and\ \bibinfo {author}
  {\bibfnamefont{D.~C.}\ \bibnamefont{Glattli}},\ }%
  \bibfield{journal}{%
  \Doi{10.1126/science.1126940}{\bibinfo {journal} {Science}}\ }%
  \textbf{\bibinfo {volume} {313}},\ \bibinfo {pages} {499} (\bibinfo {year}
  {2006})%
  \bibAnnoteFile{NoStop}{gabelli2006}%
\bibitem{feve2007}%
  \BibitemOpen
  \bibfield{author}{%
  \bibinfo {author} {\bibfnamefont{G.}~\bibnamefont{F\`eve}} \emph{et~al.},\ }%
  \bibfield{journal}{%
  \Doi{10.1126/science.1141243}{\bibinfo {journal} {Science}}\ }%
  \textbf{\bibinfo {volume} {316}},\ \bibinfo {pages} {1169} (\bibinfo {year}
  {2007})%
  \bibAnnoteFile{NoStop}{feve2007}%
\bibitem{mahe2010}%
  \BibitemOpen
  \bibfield{author}{%
  \bibinfo {author} {\bibfnamefont{A.}~\bibnamefont{Mah\'e}} \emph{et~al.},\ }%
  \bibfield{journal}{%
  \Doi{10.1103/PhysRevB.82.201309}{\bibinfo {journal} {Phys. Rev. B}}\ }%
  \textbf{\bibinfo {volume} {82}},\ \bibinfo {pages} {201309} (\bibinfo {year}
  {2010})%
  \bibAnnoteFile{NoStop}{mahe2010}%
\bibitem{buttiker1993}%
  \BibitemOpen
  \bibfield{author}{%
  \bibinfo {author} {\bibfnamefont{M.}~\bibnamefont{B\"uttiker}}, \bibinfo
  {author} {\bibfnamefont{A.}~\bibnamefont{Pr\^etre}},\ and\ \bibinfo {author}
  {\bibfnamefont{H.}~\bibnamefont{Thomas}},\ }%
  \bibfield{journal}{%
  \Doi{10.1103/PhysRevLett.70.4114}{\bibinfo {journal} {Phys. Rev. Lett.}}\ }%
  \textbf{\bibinfo {volume} {70}},\ \bibinfo {pages} {4114} (\bibinfo {year}
  {1993})%
  \bibAnnoteFile{NoStop}{buttiker1993}%
\bibitem{buttiker1993b}%
  \BibitemOpen
  \bibfield{author}{%
  \bibinfo {author} {\bibfnamefont{M.}~\bibnamefont{B\"uttiker}}, \bibinfo
  {author} {\bibfnamefont{H.}~\bibnamefont{Thomas}},\ and\ \bibinfo {author}
  {\bibfnamefont{A.}~\bibnamefont{Pr\^etre}},\ }%
  \bibfield{journal}{%
  \bibinfo {journal} {Phys. Lett. A}\ }%
  \textbf{\bibinfo {volume} {180}},\ \bibinfo {pages} {364} (\bibinfo {year}
  {1993})%
  \bibAnnoteFile{NoStop}{buttiker1993b}%
\bibitem{nigg2006}%
  \BibitemOpen
  \bibfield{author}{%
  \bibinfo {author} {\bibfnamefont{S.~E.}\ \bibnamefont{Nigg}}, \bibinfo
  {author} {\bibfnamefont{R.}~\bibnamefont{L\'opez}},\ and\ \bibinfo {author}
  {\bibfnamefont{M.}~\bibnamefont{B\"uttiker}},\ }%
  \bibfield{journal}{%
  \Doi{10.1103/PhysRevLett.97.206804}{\bibinfo {journal} {Phys. Rev. Lett.}}\
  }%
  \textbf{\bibinfo {volume} {97}},\ \bibinfo {pages} {206804} (\bibinfo {year}
  {2006})%
  \bibAnnoteFile{NoStop}{nigg2006}%
\bibitem{mora2010}%
  \BibitemOpen
  \bibfield{author}{%
  \bibinfo {author} {\bibfnamefont{C.}~\bibnamefont{Mora}}\ and\ \bibinfo
  {author} {\bibfnamefont{K.}~\bibnamefont{Le~Hur}},\ }%
  \bibfield{journal}{%
  \bibinfo {journal} {Nature Phys.}\ }%
  \textbf{\bibinfo {volume} {6}},\ \bibinfo {pages} {697} (\bibinfo {year}
  {2010})%
  \bibAnnoteFile{NoStop}{mora2010}%
\bibitem{hamamoto2010}%
  \BibitemOpen
  \bibfield{author}{%
  \bibinfo {author} {\bibfnamefont{Y.}~\bibnamefont{Hamamoto}}, \bibinfo
  {author} {\bibfnamefont{T.}~\bibnamefont{Jonckheere}}, \bibinfo {author}
  {\bibfnamefont{T.}~\bibnamefont{Kato}},\ and\ \bibinfo {author}
  {\bibfnamefont{T.}~\bibnamefont{Martin}},\ }%
  \bibfield{journal}{%
  \Doi{10.1103/PhysRevB.81.153305}{\bibinfo {journal} {Phys. Rev. B}}\ }%
  \textbf{\bibinfo {volume} {81}},\ \bibinfo {pages} {153305} (\bibinfo {year}
  {2010})%
  \bibAnnoteFile{NoStop}{hamamoto2010}%
\bibitem{hewson1997}%
  \BibitemOpen
  \bibfield{author}{%
  \bibinfo {author} {\bibfnamefont{A.~C.}\ \bibnamefont{Hewson}},\ }%
  \emph{\bibinfo {title} {The Kondo Problem to Heavy Fermions}}\ (\bibinfo
  {publisher} {Cambridge University Press, Cambridge},\ \bibinfo {year}
  {1993})%
  \bibAnnoteFile{NoStop}{hewson1997}%
\bibitem{delattre2009}%
  \BibitemOpen
  \bibfield{author}{%
  \bibinfo {author} {\bibfnamefont{T.}~\bibnamefont{Delattre}}, \bibinfo
  {author} {\bibfnamefont{C.}~\bibnamefont{Feuillet-Palma}}, \bibinfo {author}
  {\bibfnamefont{L.~G.}\ \bibnamefont{Herrmann}}, \emph{et~al.},\ }%
  \bibfield{journal}{%
  \bibinfo {journal} {Nature Phys.}\ }%
  \textbf{\bibinfo {volume} {5}},\ \bibinfo {pages} {208} (\bibinfo {year}
  {2009})%
  \bibAnnoteFile{NoStop}{delattre2009}%
\bibitem{roch2008}%
  \BibitemOpen
  \bibfield{author}{%
  \bibinfo {author} {\bibfnamefont{N.}~\bibnamefont{Roch}}, \bibinfo {author}
  {\bibfnamefont{S.}~\bibnamefont{Florens}}, \bibinfo {author}
  {\bibfnamefont{V.}~\bibnamefont{Bouchiat}}, \bibinfo {author}
  {\bibfnamefont{W.}~\bibnamefont{Wernsdorfer}},\ and\ \bibinfo {author}
  {\bibfnamefont{F.}~\bibnamefont{Balestro}},\ }%
  \bibfield{journal}{%
  \bibinfo {journal} {Nature}\ }%
  \textbf{\bibinfo {volume} {453}},\ \bibinfo {pages} {633} (\bibinfo {year}
  {2008})%
  \bibAnnoteFile{NoStop}{roch2008}%
\bibitem{lee2011}%
  \BibitemOpen
  \bibfield{author}{%
  \bibinfo {author} {\bibfnamefont{M.}~\bibnamefont{Lee}}, \bibinfo {author}
  {\bibfnamefont{R.}~\bibnamefont{L\'opez}}, \bibinfo {author}
  {\bibfnamefont{M.-S.}\ \bibnamefont{Choi}}, \bibinfo {author}
  {\bibfnamefont{T.}~\bibnamefont{Jonckheere}},\ and\ \bibinfo {author}
  {\bibfnamefont{T.}~\bibnamefont{Martin}},\ }%
  \bibfield{journal}{%
  \Doi{10.1103/PhysRevB.83.201304}{\bibinfo {journal} {Phys. Rev. B}}\ }%
  \textbf{\bibinfo {volume} {83}},\ \bibinfo {pages} {201304} (\bibinfo {year}
  {2011})%
  \bibAnnoteFile{NoStop}{lee2011}%
\bibitem{rodionov2009}%
  \BibitemOpen
  \bibfield{author}{%
  \bibinfo {author} {\bibfnamefont{Y.~I.}\ \bibnamefont{Rodionov}}, \bibinfo
  {author} {\bibfnamefont{I.~S.}\ \bibnamefont{Burmistrov}},\ and\ \bibinfo
  {author} {\bibfnamefont{A.~S.}\ \bibnamefont{Ioselevich}},\ }%
  \bibfield{journal}{%
  \Doi{10.1103/PhysRevB.80.035332}{\bibinfo {journal} {Phys. Rev. B}}\ }%
  \textbf{\bibinfo {volume} {80}},\ \bibinfo {pages} {035332} (\bibinfo {year}
  {2009})%
  \bibAnnoteFile{NoStop}{rodionov2009}%
\bibitem{nozieres1974}%
  \BibitemOpen
  \bibfield{author}{%
  \bibinfo {author} {\bibfnamefont{P.}~\bibnamefont{Nozi\`eres}},\ }%
  \bibfield{journal}{%
  \bibinfo {journal} {J. Low Temp. Phys.}\ }%
  \textbf{\bibinfo {volume} {17}},\ \bibinfo {pages} {31} (\bibinfo {year}
  {1974})%
  \bibAnnoteFile{NoStop}{nozieres1974}%
\bibitem{krishna1980a}%
  \BibitemOpen
  \bibfield{author}{%
  \bibinfo {author} {\bibfnamefont{H.~R.}\ \bibnamefont{Krishna-murthy}},
  \bibinfo {author} {\bibfnamefont{J.~W.}\ \bibnamefont{Wilkins}},\ and\
  \bibinfo {author} {\bibfnamefont{K.~G.}\ \bibnamefont{Wilson}},\ }%
  \bibfield{journal}{%
  \Doi{10.1103/PhysRevB.21.1003}{\bibinfo {journal} {Phys. Rev. B}}\ }%
  \textbf{\bibinfo {volume} {21}},\ \bibinfo {pages} {1003} (\bibinfo {year}
  {1980})%
  \bibAnnoteFile{NoStop}{krishna1980a}%
\bibitem{krishna1980b}%
  \BibitemOpen
  \bibfield{author}{%
  \bibinfo {author} {\bibfnamefont{H.~R.}\ \bibnamefont{Krishna-murthy}},
  \bibinfo {author} {\bibfnamefont{J.~W.}\ \bibnamefont{Wilkins}},\ and\
  \bibinfo {author} {\bibfnamefont{K.~G.}\ \bibnamefont{Wilson}},\ }%
  \bibfield{journal}{%
  \Doi{10.1103/PhysRevB.21.1044}{\bibinfo {journal} {Phys. Rev. B}}\ }%
  \textbf{\bibinfo {volume} {21}},\ \bibinfo {pages} {1044} (\bibinfo {year}
  {1980})%
  \bibAnnoteFile{NoStop}{krishna1980b}%
\bibitem{shiba1975}%
  \BibitemOpen
  \bibfield{author}{%
  \bibinfo {author} {\bibfnamefont{H.}~\bibnamefont{Shiba}},\ }%
  \bibfield{journal}{%
  \bibinfo {journal} {Prog. Theor. Phys.}\ }%
  \textbf{\bibinfo {volume} {54}},\ \bibinfo {pages} {967} (\bibinfo {year}
  {1975})%
  \bibAnnoteFile{NoStop}{shiba1975}%
\bibitem{tsvelick1983}%
  \BibitemOpen
  \bibfield{author}{%
  \bibinfo {author} {\bibfnamefont{A.~M.}\ \bibnamefont{Tsvelick}}\ and\
  \bibinfo {author} {\bibfnamefont{P.~B.}\ \bibnamefont{Wiegmann}},\ }%
  \bibfield{journal}{%
  \bibinfo {journal} {Adv. Phys.}\ }%
  \textbf{\bibinfo {volume} {32}},\ \bibinfo {pages} {453} (\bibinfo {year}
  {1983})%
  \bibAnnoteFile{NoStop}{tsvelick1983}%
\bibitem{kawakami1982}%
  \BibitemOpen
  \bibfield{author}{%
  \bibinfo {author} {\bibfnamefont{N.}~\bibnamefont{Kawakami}}\ and\ \bibinfo
  {author} {\bibfnamefont{A.}~\bibnamefont{Okiji}},\ }%
  \bibfield{journal}{%
  \bibinfo {journal} {J. Phys. Soc. Jpn}\ }%
  \textbf{\bibinfo {volume} {51}},\ \bibinfo {pages} {1145} (\bibinfo {year}
  {1982})%
  \bibAnnoteFile{NoStop}{kawakami1982}%
\bibitem{okiji1982}%
  \BibitemOpen
  \bibfield{author}{%
  \bibinfo {author} {\bibfnamefont{A.}~\bibnamefont{Okiji}}\ and\ \bibinfo
  {author} {\bibfnamefont{N.}~\bibnamefont{Kawakami}},\ }%
  \bibfield{journal}{%
  \bibinfo {journal} {J. Phys. Soc. Jpn}\ }%
  \textbf{\bibinfo {volume} {51}},\ \bibinfo {pages} {3192} (\bibinfo {year}
  {1982})%
  \bibAnnoteFile{NoStop}{okiji1982}%
\bibitem{haldane1978}%
  \BibitemOpen
  \bibfield{author}{%
  \bibinfo {author} {\bibfnamefont{F.~D.~M.}\ \bibnamefont{Haldane}},\ }%
  \bibfield{journal}{%
  \Doi{10.1103/PhysRevLett.40.416}{\bibinfo {journal} {Phys. Rev. Lett.}}\ }%
  \textbf{\bibinfo {volume} {40}},\ \bibinfo {pages} {416} (\bibinfo {year}
  {1978})%
  \bibAnnoteFile{NoStop}{haldane1978}%
\bibitem{garst2005}%
  \BibitemOpen
  \bibfield{author}{%
  \bibinfo {author} {\bibfnamefont{M.}~\bibnamefont{Garst}}, \bibinfo {author}
  {\bibfnamefont{P.}~\bibnamefont{W\"olfle}}, \bibinfo {author}
  {\bibfnamefont{L.}~\bibnamefont{Borda}}, \bibinfo {author}
  {\bibfnamefont{J.}~\bibnamefont{von Delft}},\ and\ \bibinfo {author}
  {\bibfnamefont{L.}~\bibnamefont{Glazman}},\ }%
  \bibfield{journal}{%
  \Doi{10.1103/PhysRevB.72.205125}{\bibinfo {journal} {Phys. Rev. B}}\ }%
  \textbf{\bibinfo {volume} {72}},\ \bibinfo {pages} {205125} (\bibinfo {year}
  {2005})%
  \bibAnnoteFile{NoStop}{garst2005}%
\bibitem{Note1}%
  \BibitemOpen
  \bibinfo {note} {The effective Hamiltonian in Eq.~\protect \textup {\hbox
  {\mathsurround \z@ \protect \normalfont (\ignorespaces \ref {fermi}\unskip
  \@@italiccorr )}} is usually accompanied by the Fermi liquid corrections
  introduced by Nozi\` eres~\cite {nozieres1974,lesage1999,*mora2009}. These
  corrections correspond however to irrelevant operators and give only
  subleading contributions to the generalized Korringa-Shiba Eq.~\protect
  \textup {\hbox {\mathsurround \z@ \protect \normalfont (\ignorespaces \ref
  {korringa}\unskip \@@italiccorr )}}.}%
  \bibAnnoteFile{Stop}{Note1}%
\bibitem{LeHur}%
  \BibitemOpen
  \bibfield{author}{%
  \bibinfo {author} {\bibfnamefont{K.}~\bibnamefont{Le~Hur}},\ }%
  \bibfield{journal}{%
  \bibinfo {journal} {Phys. Rev. Lett.}\ }%
  \textbf{\bibinfo {volume} {92}},\ \bibinfo {pages} {196804} (\bibinfo {year}
  {2004})%
  \bibAnnoteFile{NoStop}{LeHur}%
\bibitem{cottet2011}%
  \BibitemOpen
  \bibfield{author}{%
  \bibinfo {author} {\bibfnamefont{A.}~\bibnamefont{Cottet}}, \bibinfo {author}
  {\bibfnamefont{C.}~\bibnamefont{Mora}},\ and\ \bibinfo {author}
  {\bibfnamefont{T.}~\bibnamefont{Kontos}},\ }%
  \bibfield{journal}{%
  \Doi{10.1103/PhysRevB.83.121311}{\bibinfo {journal} {Phys. Rev. B}}\ }%
  \textbf{\bibinfo {volume} {83}},\ \bibinfo {pages} {121311} (\bibinfo {year}
  {2011})%
  \bibAnnoteFile{NoStop}{cottet2011}%
\bibitem{kawakami1990}%
  \BibitemOpen
  \bibfield{author}{%
  \bibinfo {author} {\bibfnamefont{N.}~\bibnamefont{Kawakami}}\ and\ \bibinfo
  {author} {\bibfnamefont{A.}~\bibnamefont{Okiji}},\ }%
  \bibfield{journal}{%
  \Doi{10.1103/PhysRevB.42.2383}{\bibinfo {journal} {Phys. Rev. B}}\ }%
  \textbf{\bibinfo {volume} {42}},\ \bibinfo {pages} {2383} (\bibinfo {year}
  {1990})%
  \bibAnnoteFile{NoStop}{kawakami1990}%
\bibitem{KLH}%
  \BibitemOpen
  \bibfield{author}{%
  \bibinfo {author} {\bibfnamefont{K.}~\bibnamefont{Le~Hur}}\ and\ \bibinfo
  {author} {\bibfnamefont{G.}~\bibnamefont{Seelig}},\ }%
  \bibfield{journal}{%
  \bibinfo {journal} {Phys. Rev. B}\ }%
  \textbf{\bibinfo {volume} {65}},\ \bibinfo {pages} {165338} (\bibinfo {year}
  {2002})%
  \bibAnnoteFile{NoStop}{KLH}%
\bibitem{etzioni2011}%
  \BibitemOpen
  \bibfield{author}{%
  \bibinfo {author} {\bibfnamefont{Y.}~\bibnamefont{Etzioni}}, \bibinfo
  {author} {\bibfnamefont{B.}~\bibnamefont{Horovitz}},\ and\ \bibinfo {author}
  {\bibfnamefont{P.}~\bibnamefont{Le~Doussal}},\ }%
  \bibfield{journal}{%
  \Doi{10.1103/PhysRevLett.106.166803}{\bibinfo {journal} {Phys. Rev. Lett.}}\
  }%
  \textbf{\bibinfo {volume} {106}},\ \bibinfo {pages} {166803} (\bibinfo {year}
  {2011})%
  \bibAnnoteFile{NoStop}{etzioni2011}%
\bibitem{lesage1999}%
  \BibitemOpen
  \bibfield{author}{%
  \bibinfo {author} {\bibfnamefont{F.}~\bibnamefont{Lesage}}\ and\ \bibinfo
  {author} {\bibfnamefont{H.}~\bibnamefont{Saleur}},\ }%
  \bibfield{journal}{%
  \Doi{10.1103/PhysRevLett.82.4540}{\bibinfo {journal} {Phys. Rev. Lett.}}\ }%
  \textbf{\bibinfo {volume} {82}},\ \bibinfo {pages} {4540} (\bibinfo {year}
  {1999})%
  \bibAnnoteFile{NoStop}{lesage1999}%
\bibitem{mora2009}%
  \BibitemOpen
  \bibfield{author}{%
  \bibinfo {author} {\bibfnamefont{C.}~\bibnamefont{Mora}},\ }%
  \bibfield{journal}{%
  \Doi{10.1103/PhysRevB.80.125304}{\bibinfo {journal} {Phys. Rev. B}}\ }%
  \textbf{\bibinfo {volume} {80}},\ \bibinfo {pages} {125304} (\bibinfo {year}
  {2009})%
  \bibAnnoteFile{NoStop}{mora2009}%
\end{thebibliography}%

\end{document}